# A VARIABLE-ENERGY CW COMPACT ACCELERATOR FOR ION CANCER THERAPY

**C. Johnstone** [*]
Fermi National Accelerator Laboratory
Fermi National Accelerator Lab, Batavia, IL 60510
cjj@fnal.gov

**J. Taylor and R. Edgecock**
Institute for Accelerator Applications
Dept. of Engineering
University of Huddersfield
Queensgate, Huddersfield, HD1 3DH
Jordan.Taylor@hud.ac.uk,T.R.Edgecock@hud.ac.uk.

**R. Schulte**
Loma Linda University
24851 Circle Drive
Loma Linda, CA 92354
RSchulte@llu.edu

**ABSTRACT**

Cancer is the second-largest cause of death in the U.S. and approximately two-thirds of all cancer patients will receive radiation therapy with the majority of the radiation treatments performed using x-rays produced by electron linacs. Charged particle beam radiation therapy, both protons and light ions, however, offers advantageous physical-dose distributions over conventional photon radiotherapy, and, for particles heavier than protons, a significant biological advantage. Despite recognition of potential advantages, there is almost no research activity in this field in the U.S. due to the lack of clinical accelerator facilities offering light ion therapy in the States. In January, 2013, a joint DOE/NCI workshop was convened to address the challenges of light ion therapy [1], inviting more than 60 experts from diverse fields related to radiation therapy. This paper reports on the conclusions of the workshop, then translates the clinical requirements into accelerator and beam-delivery technical specifications. A comparison of available or feasible accelerator technologies is compared, including a new concept for a compact, CW, and variable energy light ion accelerator currently under development. This new light ion accelerator is based on advances in nonscaling Fixed-Field Alternating gradient (FFAG) accelerator design. The new design concepts combine isochronous orbits with long (up to 4m) straight sections in a compact racetrack format allowing inner circulating orbits to be energy selected for low-loss, CW extraction, effectively eliminating the high-loss energy degrader in conventional CW cyclotron designs.

[*] Work supported by Fermi Research Alliance, LLC under contract No. DE-AC02-07CH11359

1. INTRODUCTION

Cancer is the second-largest cause of death in the U.S. About half of all cancer patients receive definitive radiation therapy either as their primary treatment or in combination with chemotherapy or surgery; overall approximately two-thirds of all cancer patients will receive radiation therapy at some point during their illness. Radiation therapy is currently a very dynamic research field driven by new technology developments. The majority of radiation treatments are still performed with linacs that generate energetic electron beams and x-rays. However particle beam therapy using proton and carbon ion beams has rapidly evolved into a new frontier in cancer therapy.

Charged particle beam radiation therapy with protons and ions offers advantageous physical-dose distributions over photon radiotherapy, and, for particles heavier than protons, a potential biological advantage. Despite recognition of reported advantages, there is almost no research activity in this field in the U.S. due to the lack of clinical accelerator facilities offering carbon and other light ion therapy in the U.S. Recognizing this problem, the NCI jointly with DOE recently organized a workshop on ion beam therapy where more than 60 experts from diverse fields related to radiation therapy were asked to define research and technical needs for advancing charged particle therapy. The detailed final report of this workshop [1] serves as a baseline resource for planning a next-generation Particle Beam Therapy Research and Development Center and is utilized as the basis for the clinical and technical specifications outlined in this paper. The workshop was followed by a NCI grant to fund planning efforts for research centers in conjunction with independent commitments to construct Particle Beam Radiation Therapy facilities in the U.S. Scientific and technology consortia involving leading institutions experienced in particle therapy and accelerator technology have formed to develop the next generation of ion beam accelerators and therapy technologies. At the workshop a list of clinical requirements were compiled in order to address outstanding treatment issues and research needs to be incorporated into the planning effort and next stage towards a facility. These include reducing range uncertainties, managing tumor and organ motion, proximity of tumors to critical normal structures, matching the biological effectiveness of different ion beams to specific tumor characteristics, clinical trials and protocols for hypfractionation and multi-ion treatments, in addition to basic radiobiology research. The clinical requirements identified in the workshop for a National Particle Beam Research Center and subsequently extracted from the report are summarized in Table I.

The cost and footprint of an ion therapy facility are considered the primary economical obstacles to widespread adoption of this promising therapy. New accelerator technologies must address capital costs in the form of compact accelerators and beam delivery systems; the latter especially in ion gantries. Compact therapy accelerators to date are represented by AVF cyclotrons with energy degraders. Higher fixed-energy ion cyclotrons have not been adopted over the more costly synchrotron due to increased shielding and radio-activation issues. Although linacs can produce rapid energy changes combined with high dose deposition rates, they remain the most costly of the accelerator systems. The solution proposed here is a new type of FFAG accelerator which combines advantageous features of the synchrotron and cyclotron: strong focusing machine tunes with low-loss variable-energy extraction and a CW beam format, respectively.

2. SPECIFICATIONS FOR A LIGHT ION FACILITY

Progress on the issues and requirements identified in the workshop necessarily rely on technical innovation in accelerator, beam delivery, and imaging systems. Some goals such as reducing range uncertainties, accurately differentiating between the tumor and normal tissues to enable precision dose delivery while minimizing the radiation dose to normal tissue can only be realized through incorporating particle-beam imaging and range measurements into the treatment planning –combined with real-time dose verification during treatment.  Imaging, particularly pCT, however, are not simple add-ons – they must be integrated into the accelerator and beam delivery at the conceptual design stage.  Switching rapidly between ion species is an integral part of a state-of-the-art facility.

**Table I. Ion Therapy Facility clinical requirements per NCI/DOE report [1].**

| **Multi-ion capability** | p, He, Li, B, C (O and Ne also desirable) <br> Fast switching between ion species (≤1 sec) |
|---|---|
| **Energy range** | 60 MeV/nucleon to 430 MeV/nucleon <br> Depths up to 30 cm for carbon ions |
| **Field size** | At least 20 x 20 cm$^2$ (optimally up to 40 x 40 cm$^2$) |
| **Real-time imaging (radiography and CT):** <br> For tumor position verification and motion control | For patient sizes up to 60 cm in depth. |
| **Dose delivery rates:** <br> **Minimum requirement** <br> **Hypofractionation treatments in under 1 minute, (ideally in one breath hold for motion control)** | 20 Gy/minute <br> up to 5-8 Gy/8 sec (breath-hold) for a cubic liter |
| **Pencil beam scanning:** <br> **Fast treatment for a large variety of tumor sizes and shapes.  Two extremes are considered:  30 cm x 30 cm tumor single layer in depth and a cubic volume** | Transverse scanning rate of 1-10 cm/msec <br> Energy step time of 10-100 msec <br> (These are present state-of-the-art for NC and SC magnetic components) |
| **Transverse beam size:** <br> **selectable, with stable, Gaussian profiles.** | 3 mm to 10 mm FWHM |
| **Energy step size** | Protons:  2 MeV (~0.25 cm in range) <br> Carbon: 2 MeV/nucleon (~0.1 cm in range) |
| **Lateral targeting accuracy at the Bragg peak** | Protons: ±0.5 mm <br> Carbon: ±0.2 mm |
| **Dose accuracy/fraction** | 2.5% monitored at ≥40 kHz during dose deposition |

**2.1 Accelerator and Beam Delivery Performance Specifications**

Requirements for high dose rates and effective motion management put challenging requirements on beam scanning rates and energy step times.  To maintain safety standards with high dose rates next-generation machines will have to rely on fast real-time imaging and improved monitoring and controls technology. Preparation for a final end-to-end facility concept must assure compatible performance at all levels of every system from multi-ion sources through gantry beam delivery. Not only must next-generation accelerators, beam-delivery, and imaging technologies meet exceptionally demanding clinical requirements listed here, civil construction and operational costs must be reduced significantly from current facilities without compromising expanded capability.

Many requirements listed in Table I entail innovative integration of diverse and advanced delivery and diagnostic development, but all systems must be supported by the accelerator beam format and delivery modes. To better understand the intersection of the accelerator and beam delivery with the clinical, the requirements in Table 1 are translated into actual dose deposition rates expressed in terms of machine and delivery performance parameters. These are presented in Table II and are then used for design, selection, and comparison purposes between accelerator systems.

In Table II, two different dose deposition criteria were applied, based on minute and second timescales, and extrapolated into estimates of beam intensity. For example, the technical specification for a proton Computed Tomography (pCT) imaging system is only 106-107 particle/sec and preferably CW due to intrinsic individual-particle tracking and reconstruction limitations. This is a very different condition and demand on the accelerator, delivery, and controls as compared with an escalated dose rate of $10^{12}$ p/sec for target motion management. Combining the two results represents an unprecedented, but achievable, dynamical range for accelerator operation. Short high dose rates on second timescales remain the most demanding environment for accelerator performance. (It should be noted that longer timescales will likely be used in practice; i.e. the breath-hold rates, but the intent here is to define upper limits on accelerator performance which could be envisioned.)

**Table II. Dose delivery rates expressed in terms of accelerator performance for two different timescales (minutes and seconds or a breath hold period of 5-8 sec) for different dose regimes: standard fractions, hypofractions, and radiobiology research. Assumes average of 109 p/cm$^2$ for a 1 Gy dose. For carbon this should be divided by the RBE ratio which is on average, ~3. These two timescales are useful for implementation of higher-dose protocols.**

| Dose Delivery Rate | 30 x 30 cm$^2$ field<br>single layer/energy<br>sweep step size 5 mm | 10 x 10 x 10 cm$^3$ field<br>40 layers/energy steps<br>sweep step size 5 mm |
|---|---|---|
| **Ions: (charge to mass ½) minimum set**<br>Additional ions - optimal | $H_2^+, {}^4He^{2+}, {}^6Li^{3+}, {}^{10}B_{10}^{5+}, {}^{12}C^{6+}$<br>${}^{14}N^{7+}, {}^{16}O^{8+}, {}^{20}Ne^{10+}$ | |
| **Imaging: 1 to 10 MHz data systems** | $10^6$-$10^7$ p/sec | $10^6$-$10^7$ p/sec |
| **Normal Fraction: 1-2 Gy/fraction**<br>1 Gy/min<br>1 Gy/sec | 1/60 x $10^{12}$ p/sec<br>1 x $10^{12}$ p/sec | 4/60 x $10^{12}$ p/sec<br>4 x $10^{12}$ p/sec* |
| **Hypofraction Regime: 5-8 Gy/fraction**<br>5-8 Gy/min<br>5-8 Gy/sec<br>5-8 Gy/breathhold | (5-8)/60 x $10^{12}$ p/sec<br>5-8 x $10^{12}$ p/sec<br>1 x $10^{12}$ p/sec | (5-8)/60 x $10^{12}$ p/sec<br>2-3 x $10^{13}$ p/sec*<br>4 x $10^{12}$ p/sec* |
| **Radiobiology: up to 20 Gy/fraction**<br>20 Gy/min<br>20 Gy/sec<br>20 Gy/breathhold | (2)/60 x $10^{13}$ p/sec<br>2 x $10^{13}$ p/sec<br>2-4 x $10^{12}$ p/sec | 8/60 x $10^{13}$ p/sec<br>8 x $10^{13}$ p/sec*<br>1-2 x $10^{13}$ p/sec* |

## 3. ACCELERATOR COMPARISONS

Current ion therapy facilities use slow-extracting synchrotrons, however, further developments are needed to reduce cost and footprint and overcome limitations on intensity and energy modulation. New, rapid-cycling synchrotrons may overcome the limitations of slow-extracting synchrotrons but retain a very large footprint and require new technical limits on controls technology and dose delivery. The ultra-short beam spill (33 and 16 millisec vs 1 sec) would eliminate much of the existing spill uniformity controls, diagnostics, and scanning and gating technologies. Proposed studies to explore the biology of hypofractionation with intense, ultra-short dose depositions (1 – 8 sec timescales) are not possible with a conventional synchrotron. Although many proton facilities use cyclotrons, scaling cyclotrons to the higher energies required for ion facilities presents many challenges particularly in degrader-based beam delivery. The cyclinac is an interesting combination of technologies however the upstream injector cyclotron is not well matched to the higher-energy linac with high beam losses projected. It has recently evolved into a "bent" linac layout with 180 degree achromatic arc to reduce the linear footprint. However, the single-pass linac structure remains the most costly of the ion accelerators. A relatively new approach, a CW-beam non-scaling FFAG accelerator combines many of the strengths of synchrotrons and cyclotrons, has recently matured to a prototyping and implementation stage, and is the solution promoted here. An assessment of accelerator options in terms of their capability to meet clinical requirements is summarized in Table III.

**Table III. Accelerator options and an assessment of performance for ion therapy**

| Accelerator Type | Size | Max Dose Rate /liter | Energy Modulation Var/ Fixed | Energy Modulation Machine Time | Energy Modulation Beam Delivery NC | Energy Modulation Beam Delivery SC | Scanning Transverse (T) Longitudinal (L) |
|---|---|---|---|---|---|---|---|
| SC Synchrotron | 20-40 m (diam) | 5 Gy/min 0.2 Gy/sec | V | 1 ms – 2 sec | 10 ms | 100 ms | T + L in acc cycle |
| RC Synchrotron | 60 m (circum) | ~x Cycle factor 1.3-20 Gy/min | V | 1 – 66 ms | 10 ms | 100 ms | L TBD T cyc to cyc |
| Compact Proton Synchrotron | 5 m (diam) | 0.075 Gy/min 0.0025 Gy/sec | V | 1 ms -2 sec | 10 ms | 100 ms | T + L In acc cycle |
| Hitachi | 8 m | 0.75 Gy/min 0.025 Gy/sec | V | | | | |
| Linac/Cyclinac | 40 -80m | Any rate | V | 1 ms | 10 ms | 100 ms | Any |
| Cyclotron | 6.3 m (diam) | 5 Gy/min 0.08 Gy/sec | F | 1 ms | 10 ms | 100 ms | T then L |
| FFAG | 4m x 6m racetrack | Any rate | V | 50 μsec | 10 ms | 100 ms | Any |

## 4. ACCELERATOR APPROACH

FFAG (Fixed-Field Alternating-Gradient) accelerators are a class of accelerators that merge the best features of cyclotrons and synchrotrons. They can deliver continuous beam in combination with low loss operation and variable energy, specifically at the high energies required for carbon therapy and the high intensities required for radiobiology. The FFAG successfully blends the fixed fields of cyclotrons with the strong focusing fields of synchrotrons to achieve the isochronous orbits of the cyclotron (at higher energies than the cyclotron), and the strong, constant tune of the synchrotron in an aggressively compact footprint. Isochronous orbits allow use of a fixed-frequency Radio-Frequency (RF) system enabling continuous beam, and maximum flexibility in real-time imaging and pencil-beam scanning - all at the high energies required for imaging and ion therapy. Constant tune further not only minimizes acceleration losses, it facilitates the insertion of long straight sections thereby promoting efficient, low-loss extraction as evidenced in synchrotron operation. Straight sections also allow for fast switching, even multi-ion injection and variable energy extraction. These FFAG design features appear unusually well matched to the requirements of a next-generation ion beam therapy system and are potentially economical, with costs comparable to the cyclotron. No FFAGs have yet been built specifically for radiation therapy but comparable operating FFAG (scaling) systems exist in Japan. We are presently designing and evaluating a a new design, presented here, of an ultra-compact, variable-energy, non-scaling, continuous wave (CW) FFAG for ion therapy.

The capabilities of FFAG accelerator designs meet specifications for the requirements for future ion beam therapy systems as identified in Tables 1 and 2. These requirements include the ability to deliver multiple light ion types (protons, deuterons, helium, carbon, possibly up to neon), short deposition times with image-guidance and real-time measurements of patient parameters, rapid beam scanning and energy modulation, with energies up to 430 MeV/nucleon, and with all components optimized for effective and safe operation. With multiport injection and extraction capability both imaging and treatment beams can be accelerated in the FFAG with appropriate energy and range requirements and rapid ion species switching or interleaving times (<sec). On board imaging beams would initiate at the accelerator and be transported through the beam delivery system or through a dedicated multiport delivery system for imaging, achieving the ultra-low currents required for imaging without difficult measures on the main treatment-beam accelerating systems.

### 4.1 A CW, Variable-energy FFAG Accelerator for Ion Therapy

We have designed a 70 - 430 MeV/nucleon FFAG in a racetrack layout that supports two opposing long straight sections for rapid acceleration and efficient injection and extraction, in a highly compact footprint (~11m x 15m for normal conducting and ~6m x 10m for superconducting arcs). One or more injectors can be utilized for multi-ion injection or rapid switching. This FFAG can accelerate ions with a charge to mass ratio of 1:2 (e.g. H2, He, B, Li, C) and is ideal for radiation therapy. The long synchrotron-like straight sections permit rapid variable energy extraction on the timescale of milliseconds eliminating the need for a substantive beam degrader. High-gradient synchrotron-like RF modules can also be exploited in the opposing straight for rapid, low-loss acceleration – this is to be compared to the lower-gradient Dee cavities used in compact cyclotrons. Figures 1 show basic concepts and dimensions including a footprint comparison with the rapid cycling synchrotron design for 430 MeV/nucleon. The energy of 330 MeV/nucleon was chosen to support predominately He therapy @250 MeV/nucleon with particle-beam imaging which requires 330 MeV/nucleon. Lower energy carbon ion therapy can also be applied with this accelerator.

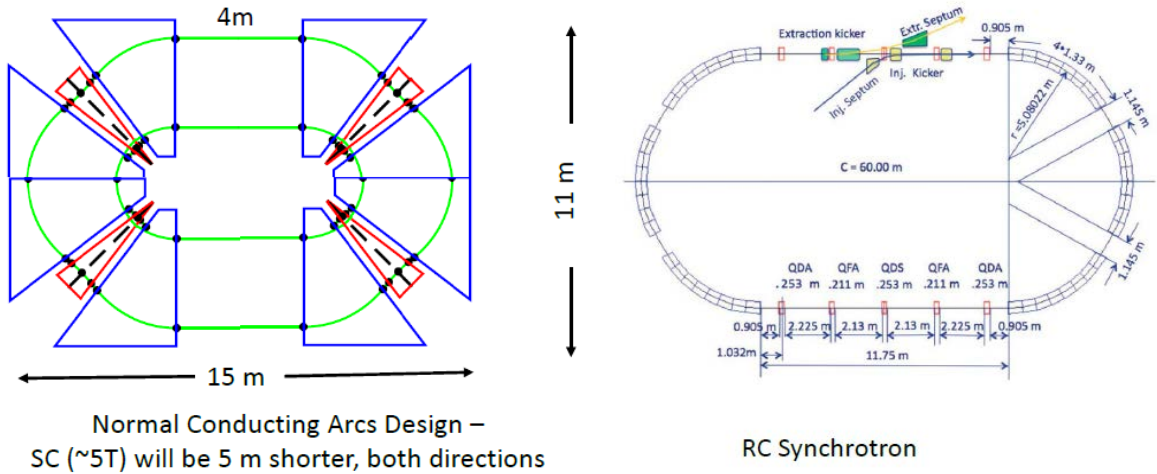

**Figure 1. Comparison of a 300 MeV/nucleon ion FFAG accelerator with the 430 MeV/nucleon Rapid Cycling racetrack synchrotron (~30 x 60 m footprint).**

Preliminary studies indicate strong tune stability over the acceleration range as shown in Figure 2. Initial isochronous orbits are at ~ a percent level, with ongoing work on refining and smoothing orbits to achieve the required stability during acceleration. However the strong focusing produces a parabolic time of flight curve which allows acceleration via the serpentine channel and lower isochronous tolerances on orbital path lengths as a function of energy.

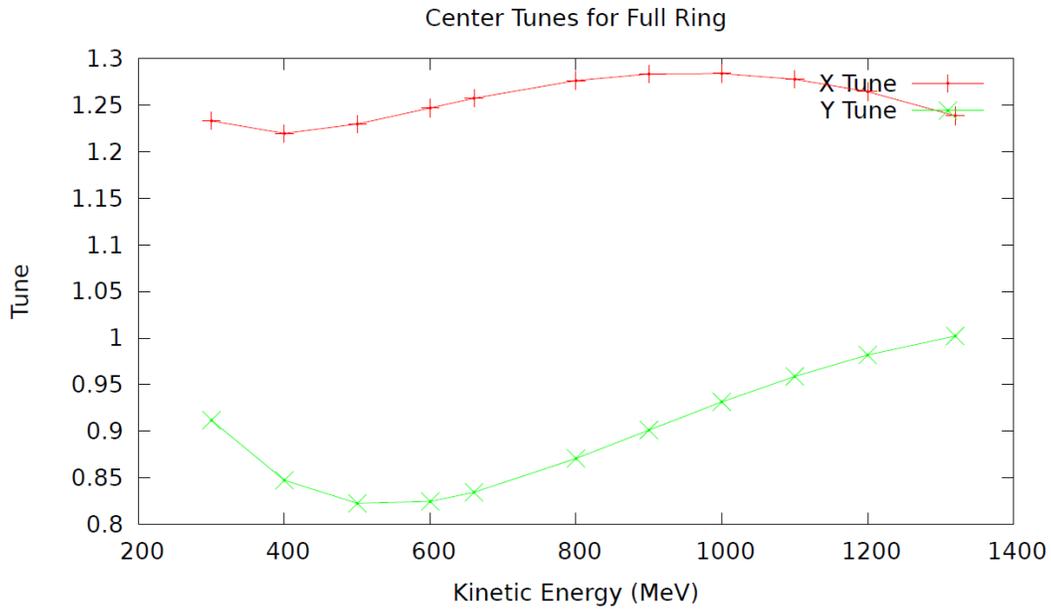

**Figure 2. Strong focusing machine tune over the 70-330 MeV/nucleon acceleration range.**

## 5. SUMMARY

Our analysis shows that a non-scaling, CW FFAG exhibits great potential as a highly compact and economically viable alternative to synchrotrons, cyclotrons, linear accelerators and other linac formats for next-generation ion beam radiation therapy. The fixed fields and fixed RF simplify operations compared to synchrotrons which must ramp fields in coordination with swept-frequency RF. The injector could be a lower-energy FFAG or cyclotron which, in principle, can serve as a standalone accelerator for certain treatments (e.g. ocular melanoma) or could be used to accelerate protons to ~300 MeV for proton Computed Tomography (pCT). This machine is currently under international development and will be ready for either a National Particle Beam Radiation Center or private ion therapy facilities on approximately a 5-year timescale.